\documentclass[fleqn,10pt]{wlscirep}
\usepackage[utf8]{inputenc}
\usepackage[T1]{fontenc}
\usepackage{braket}
\usepackage{amsmath}
\usepackage{caption}
\usepackage{subcaption}
\usepackage[normalem]{ulem}
\usepackage{multirow}

\title{Resource-efficient utilization of quantum computers}

\author[1]{Ijaz Ahamed Mohammad}
\author[1,2]{Matej Pivoluska}
\author[1,3]{Martin Plesch*}
\affil[1]{Institute of Physics, Slovak Academy of Sciences, Dúbravská cesta 9, 841 04 Bratislava, Slovak Republic}
\affil[2]{Institute of Computer Science, Masaryk University, Šumavská 416, 602 00 Brno, Czech Republic}
\affil[3]{Faculty of Natural sciences and Informatics, Constantine the Philosopher University, Tr. A. Hlinku 1, 949 01 Nitra, Slovakia}
\affil[*]{martin.plesch@savba.sk}

\begin{abstract}
The current state of quantum computing is commonly described as the Noisy Intermediate-Scale Quantum era. Available computers contain a few dozens of qubits and can perform a few dozens of operations before the inevitable noise erases all information encoded in the calculation. Even if the technology advances fast within the next years, any use of quantum computers will be limited to short and simple tasks, serving as subroutines of more complex classical procedures. Even for these applications the resource efficiency, measured in the number of quantum computer runs, will be a key parameter. Here we suggest a general optimization procedure for hybrid quantum-classical algorithms that allows finding the optimal approach with limited quantum resources. We demonstrate this procedure on a specific example of variational quantum algorithm used to find the ground state energy of a hydrogen molecule. 

\end{abstract}
\begin{document}

\flushbottom
\maketitle
% * <john.hammersley@gmail.com> 2015-02-09T12:07:31.197Z:
%
%  Click the title above to edit the author information and abstract
%
\thispagestyle{empty}

\section*{Introduction}

Quantum computers, as a theoretical concept, have been suggested in the 1980’s independently by Paul Benioff \cite{Benioff1980} and Yuri Manin \cite{manin_computable_1980}. Later they have been popularized by Richard Feynman in his seminal work  on simulating quantum physics with a quantum mechanical computer \cite{Feynman1982}. 
In the following $30$ years the potential of quantum computers to outperform classical computers in certain tasks was thoroughly studied \cite{nielsen00} and many important quantum algorithms have been found, among them an algorithm to factorize large numbers in polynomial time \cite{shor} and an algorithm for a fast search in unstructured databases \cite{grover}.
In recent years experimental quantum computing has achieved tremendous advances and thus the design of quantum algorithms has shifted from purely theoretical research towards more practical questions.
One of the main new areas of research is the practical utilization of contemporary quantum processors, which are encumbered by noise that is reducing their reliability. 
These are called Noisy Intermediate-Scale Quantum (NISQ) computers \cite{Preskill2018quantumcomputingin} and in attempt to utilize them efficiently many so-called variational algorithms have been designed \cite{Cerezo2021,RevModPhys.94.015004}. 
Variational algorithms use both quantum and classical computational resources and their main idea is to design a parameterized quantum circuit and a measurement method associated to a minimization problem at hand.
The parameters of the quantum circuit minimizing the value of the target function are then searched for using classical function minimization subroutines.
Algorithms of this type can be designed for optimization problems in a plethora of real world use cases, ranging from chemistry \cite{mccaskey2019quantum,mcardle2020quantum}, through artificial intelligence \cite{dunjko2018machine,biamonte2017quantum} to financial market modelling \cite{orus2019quantum}.

Such a combination of classical and quantum approaches leads to many interesting challenges. The accuracy and reliability of the final output is inevitably limited by a combination of different factors. First, the classical optimization algorithm is not guaranteed to converge to the true minimum. 
This is because many optimization algorithms are stochastic by nature and they use randomness at least during initialization to choose the starting point. In many scenarios, randomness is required in each iteration. This randomness ensures that the optimization procedures can deal with a large family of different functions without getting stuck in a local minimum, but naturally also leads to stochasticity of the whole calculation, i.e. the algorithms are not guaranteed to succeed. Thus for obtaining a useful result with high probability, the procedure needs to be repeated several times. 

Secondly, even in the noiseless quantum processor scenario the outcome of any useful quantum computation is stochastic.
Typically, the outcome is a non-computational basis state (otherwise the computation would be classically efficiently simulable) and is characterized by frequencies of outcomes for different measurement settings. 
A single run of a quantum computer  only provides a single snapshot of the state for one measurement setting.
Any optimization procedure therefore inevitably comes with a trade-off between a more precise measurement on a single position in the parameter-space and less precise measurements of many positions. 
It is non-trivial to decide which of these two strategies leads to better results.

Last, but not least, one cannot rely on any kind of error correcting procedures for NISQ computers. The classical procedure itself will have to account for the fact that any result of the quantum computer is influenced, on top of the statistical fluctuations due to the intrinsic quantum randomness, also by experimental errors. All these three complications suggest that the classical procedure will have to go significantly further than just utilizing known optimization methods. 
For any near-future hybrid algorithms the main limiting factor will be the quantum part. Both from the time (queues) and financial (pay-per-shot) perspective, the efficiency of the algorithm will be measured by its ability to produce good results with as little utilization of quantum computers as possible. This generic statement can be simplified into a measure relating the reliability and precision of the result (the probability to get a result and how good the result is) to the number of utilizations of a quantum computer with a single measurement outcome (a shot). Simply speaking, the aim is to get the best possible result with a fixed budget counted in the number of shots (i.e., time or money). 

Instead of suggesting a specific new technique on the classical level, we introduce a meta-optimization technique that can be utilized for a broad variety of classical-quantum scenarios. For any classical optimization procedure and connected quantum calculation the technique first samples the fluctuation of the results depending on the parameters, estimates this dependence and suggests the optimal setting. These will be expressed in the parameters of the classical and quantum part of the procedure and the optimal number of repetitions. 
We demonstrate this approach on a specific example on a Variational Quantum Eigensolver (VQE) \cite{peruzzo2014variational} determining the ground energy of Hydrogen molecule in combination with the Simultaneous Perturbation Stochastic Approximation (SPSA) \cite{spall1998overview} classical minimization procedure. 
It turns out that for reasonable parameters, namely the number of available quantum shots counting in a few millions and the expected accuracy of the energy in small multiples of chemical precision, the optimal number of repetitions of the whole procedure is less than a dozen and the probability of obtaining the result within the expected precision ranges from $10$ \% to almost $90$ \%.  This shows that one needs to be very careful by choosing the exact strategy during the optimization process.  

The paper is organized as follows. In the rest of this chapter we introduce the VQE and SPSA. In the Results section, we describe the procedure in general terms, apply it to the specific Hydrogen molecule scenario and present the results. In Methods we provide details on the implementation of the quantum part of the algorithm on IBM quantum machines.

\subsection*{Variational Quantum Eigensolver}

\begin{figure}
    \centering
    \includegraphics[scale = 0.5]{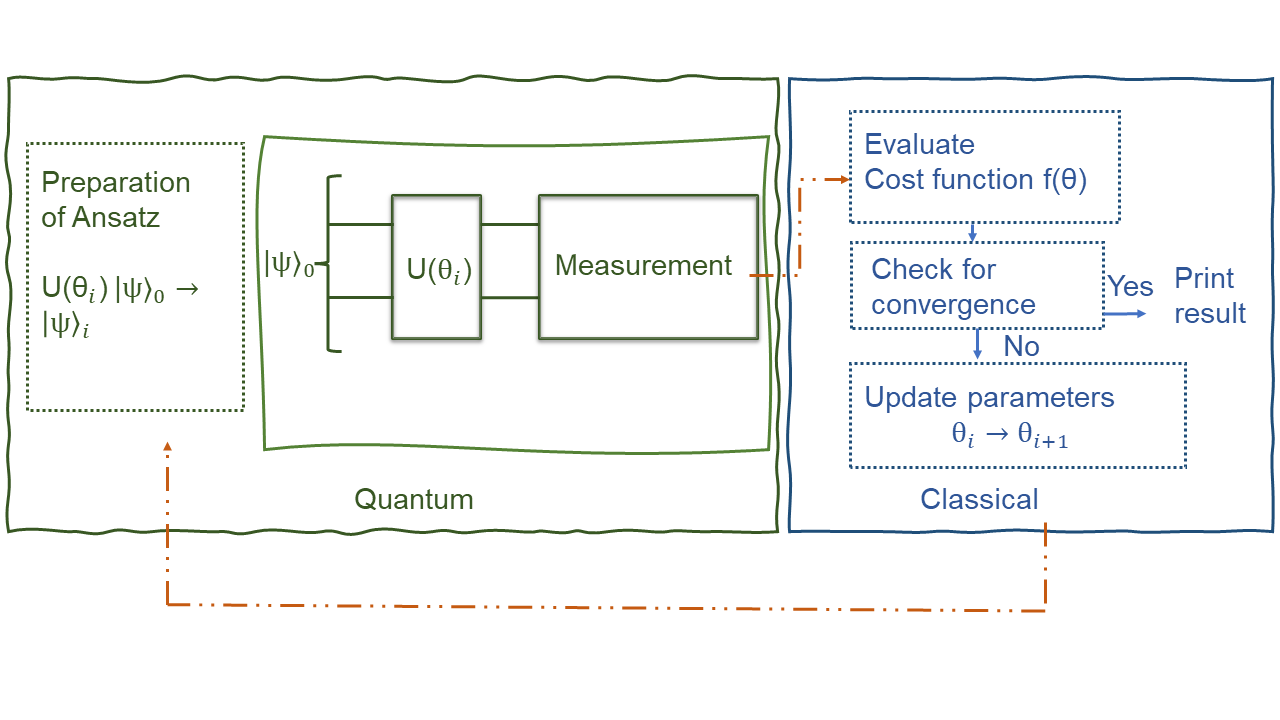}
    \caption{The VQE algorithm is divided into two parts. First, the ansatz ($\ket{\psi_{i}}$) is prepared from an initial state ($\ket{\psi_{0}}$) with the help of some parameterized circuit which can viewed as the action of a parameterized unitary operator ($U(\theta_i)$). The state is later measured and by using the counts of the outcomes of the measurement, the cost function is evaluated. If the convergence of the optimizer isn't achieved, this process is repeated until desired result with good convergence is obtained in the end.}
    \label{fig:VQE}
\end{figure}

In this paper we study variational quantum eigensolver, introduced by Peruzzo \textit{et al.} \cite{peruzzo2014variational}, and subsequently thoroughly studied 
by many authors (see review papers \cite{Cerezo2021,RevModPhys.94.015004,https://doi.org/10.48550/arxiv.2111.05176} on the topic of VQE and the references therein).
 VQE is used to calculate the ground state energy ($E_{0}$) of any Hamiltonian ($H$). It is based on quantum physics' variational principle, which asserts that any Hamiltonian's expectation with regard to any state
($\ket{\psi}$) is higher than the ground state energy:
\begin{equation}
    \braket{\psi|H|\psi} \geq E_{0}. \label{eq1}
\end{equation}
Classical techniques become impractical to detect the ground energies of the Hamiltonian as the Hamiltonian grows in size. VQE is a hybrid method that circumvents this issue by combining quantum processors with classical optimizers. The working of a VQE algorithm is as follows:
\begin{itemize}
    \item Preparation of a trial state ($\ket{\psi_{i}}$) from an initial state ($\ket{\psi_{0}}$) (typically a zero state) by introducing some parameters ($\theta_{j}$), which can be updated to change the trial state in each iteration;
    \item Measurement of the state (possibly in several bases), obtaining frequencies approximating probabilities. In this stage the trade-off is being made between a quick/cheap imprecise value and slower/more expensive precise value;
    \item Evaluation of the cost function (energy $E_{i}$) based on the probabilities approximated in the previous step;
    \item Using this evaluation as an input to the classical optimizer. This checks for convergence, and if the convergence criteria are not met, it modifies the parameters, resulting in a new iteration. This procedure is repeated until the convergence criteria are met.
\end{itemize}
The VQE algorithm's operation is depicted in Fig.~\ref{fig:VQE}.

\subsection*{Classical optimizers}

Previous research in modifying the VQE algorithm to be more resource efficient approached the problem by tailoring the classical optimization part of the algorithm to economically distribute the shot budget among various required calculations. 
This is a very fruitful approach and many significant advances have been achieved.
One example of such an algorithm is individual Coupled Adaptive Number of Shots (iCANS) \cite{Kubler2020adaptiveoptimizer}. 
The iCANS algorithm distributes the shot budget available for each iteration across all parameters in order to find partial derivatives in all directions. 
For each direction the amount of shots used to evaluate the gradient in each iteration $i$ is inversely proportional to the square of the gradient in this direction in iteration $i-1$. In such a way, low gradient directions in each iteration are evaluated with larger precision, which is necessary for successful convergence.
Another approach is to distribute the shot budget for each iteration unevenly across different Hamiltonian terms. 
This has been explored in conjunction with several optimization algorithms, with the best performance observed with iCANS algorithm resulting in an algorithm called Random Operator Sampling for Adaptive Learning with Individual Number of shots (Rosalin) \cite{rosalin}.
Last but not least, a more general discussion about evaluating only a randomly chosen subset of all possible Hamiltonian terms in each iteration, with conjunction with evaluating gradients only in a randomly chosen subset of directions (so-called stochastic gradient descent) can be found in \cite{Sweke2020stochasticgradient}.

In the commonly used stochastic gradient descent algorithms at each iteration, the partial gradient is evaluated by perturbing each parameter along and opposite to its direction requiring two function evaluations (alternatively in VQE one can use the parameter shift rule \cite{schuld2019evaluating}).
This can be computationally expensive for functions with many parameters, which suggests that we will need $2pt$ functions evaluations in total, where $p$ is the length of the parameter vector $\theta$ and $t$ is the total number of iterations used in optimization. 
In contrast, in Simultaneous Perturbation Stochastic Approximation (SPSA)\cite{spall1998overview}, the gradient in each iteration is estimated by simultaneously perturbing all parameters $\theta$ randomly with only two function evaluations required for the whole gradient. 
Thereby resulting in $2t'$ function evaluations in total where $t'$ is the number of iterations SPSA took to converge. The second advantage of SPSA is its stochasticity, which makes it noise-resistant. 
Because each parameter is already randomly perturbed, subsequent noise perturbations are less likely to disrupt the optimization process.

\subsection*{Costs of final estimation}

While rather counter-intuitive, advanced optimization techniques are able to find the parameters of the optimum even if very few shots are used per each measurement. 
In other words, one can find the target even if almost blind, with one reservation -- the resulting value of the cost function is very imprecise.
In many cases this issue is not addressed at all and the final value is calculated by classical means \cite{Kubler2020adaptiveoptimizer,rosalin,tamiya2022stochastic}. 
This, however, is a simplification -- classical recalculation of the energies becomes infeasible in instances where the quantum computer is really useful.
Thus, if one wants to have a complete solution, it turns out that at the end of the calculation one needs to test the resulting state for the precise value of the cost function, investing a large number of shots -- as we will show in the result section, this can be comparable to the whole optimization procedure budget.
This further complicates the optimization -- not only the number of repetitions of the procedure needs to be optimized, but also the distribution of the shots between the optimization procedure itself and the final estimation of the cost function value. 

\subsection*{$H_{2}$ molecule with SPSA optimizer}
We will apply the suggested method onto a specific example of Variational Quantum Eigensolver seeking the ground energy of an $H_{2}$ molecule. We considered a qubit Hamiltonian of $H_{2}$ molecule with 2-qubits with the distance between the atoms set as 0.725 \AA \  of the form 
    \begin{equation}
    H = c_{0} (I \otimes I) + c_{1} (Z\otimes I) + c_{1} (I \otimes Z) + c_{2} (Z \otimes Z) + c_{3} (X \otimes X), \label{h2}
    \end{equation}
    where $X,Z$ are the usual notations of the Pauli matrices and the coefficients $c_{0}$, $c_{1}$, $c_{2}$, $c_{3}$ are specified as $-$ 1.05016, 0.40421, 0.01135, and 0.18038 respectively\cite{mihalikova2022cost}.
Even for such a simple physical system and form of the Hamiltonian a significant cost is required to get the desired results from conventional use of the VQE algorithm. We will show by using SPSA method how our optimization can help to achieve better results.

\section*{Results}
%\guidelines{Up to three levels of \textbf{subheading} are permitted. Subheadings should not be numbered.}

In the most general terms, we suggest a meta optimization procedure with following inputs:
\begin{itemize}
\item An optimization technique parameterized solely by the number of shots used per the whole run; 
\item Final result estimation method parameterized by its costs measured in the number of shots;  
\item Shot budget, i.e. the total number of shots available;
\item Desired accuracy, i.e. how far away from the exact value of the cost function in optimum the result is considered as successful.
\end{itemize}

Regarding the last condition it is important to stress that the success of the optimization is evaluated in terms of the cost function and not the parameters.
Finding a local minimum far away from the global one (e.g. an excited state), which has the value of cost function within the desired accuracy (e.g. if the accuracy is set as low as the first energy gap) is taken as a successful result. 

The output of the meta optimization consists of
\begin{itemize}
\item Optimal number of repetitions of the optimization; 
\item Number of shots to be used per one optimization run; 
\item Number of shots to be used per final estimation;
\item Probability of success. 
\end{itemize}

Using these results, one has to take the steps as suggested, i.e., run the optimization method desired number of times, perform the final estimation of all results and select the best one based on this estimation (this might not be the best result according to the optimization method itself, due to a larger variance of energy calculations used during the run). 
The probability of success expresses the likelihood we get the result (at least) as good as we wanted. 
For large budget of shots, resulting into reliable final estimations, we will know how good the result is with high probability after obtaining it.
This is not completely the case for small budgets, resulting into unreliable final estimation -- here we might have a good result even if not estimated so, or overestimate the quality of a bad result.

Alternatively, one can re-formulate the procedure by exchanging the Success probability and Accuracy. If we state the expected probability to get a result, the meta-optimization procedure will suggest the parameters to obtain the best result (highest accuracy) with at least the given probability.

\subsection*{The cost of final estimation}\label{cofc}
%\mpi{I would introduce the existing work and use it to define the probability of obtaining the result within an absolute error.}{}

While the input-output relation for  optimization techniques are complex, the final estimation precision is in a simple mathematical relationship with the number of shots used. 
In VQAs, the objective function, which is typically an expectation of summations of different quantum operators with different weights, is calculated from the probability estimates from the quantum measurements. 
Such a quantum operator is of the form ${F}$ = $\sum_{i} A_{i}$, with ${A_i}$ denoting a collection of simultaneously-measurable quantum operators with $ {A_i} = \sum_{j} c_{j} O_{j}$, where $c_j$ and ${O_j}$ denote coefficients and Pauli words, respectively. 
The cost of a final estimate ($S$) for a desired accuracy of $\varepsilon$ is given by  \cite{yen2021cartan}
\begin{equation}
    S := \frac{\big(\sum_{i} \sqrt{\operatorname{Var}({A}_{i})}\big)^{2}}{\varepsilon^{2}}.
\end{equation}
The authors of the article \cite{yen2022deterministic} have shown that the variance can be computed as
\begin{equation}
    \operatorname{Var}(A_i) = \sum_{jk} c_{j}c_{k}\operatorname{Cov}(O_{j},O_{k}),
\end{equation}
where $\operatorname{Cov}(\cdot)$ represents the covariance between the Pauli words. As discussed in the article \cite{gonthier2022measurements}, from the Cauchy-Schawrtz inequality the covariance between two Pauli words is upper bounded as 
\begin{equation}
    \operatorname{Cov}(O_{j},O_{k}) \leq \left|\sqrt{\operatorname{Var}(O_{j})\operatorname{Var}(O_{k})}\right|. \label{eq-cov}
\end{equation}
As the magnitude of the covariance is upper bounded as shown in Eq. \ref{eq-cov}, one can consider the average from a random distribution of the covariance of cross terms to be zero. Using this and coupling it with the fact that the variance of a Pauli word is upper bounded by 1, the number of shots required to obtain the estimate of the objective function with accuracy $\varepsilon$ is given by
\begin{equation}
    S =  \frac{\left(\sum_{i} \sqrt{\sum_{j} c_{ij}^{2}}\right)^{2}}{\varepsilon^{2}}. \label{eq-m}
\end{equation}

\subsubsection*{Example of $H_2$ molecule}
Measuring the eigenstate of our Hamiltonian's ground state energy allows us to examine the impact of the statistical disturbance brought on by measurements. 
Given that the Hamiltonian matrix in our example is small, we performed a classical calculation to identify the Hamiltonian's corresponding eigenvalues and eigenstates. 
We discovered that the energy of the ground state for our Hamiltonian was $1.8671$~Ha. 
We then measured the quantum state with various numbers of shots, initialized it as the eigenstate corresponding to the ground state energy in qiskit, and repeated the experiment $10,000$ times. 
We select various desirable ranges surrounding the true value, ranging from chemical precision ($\pm 0.0015$~Ha) to $5$ times the chemical precision range ($\pm 0.0075$~Ha). 

\begin{figure}
     \centering
     \includegraphics[scale = 0.6]{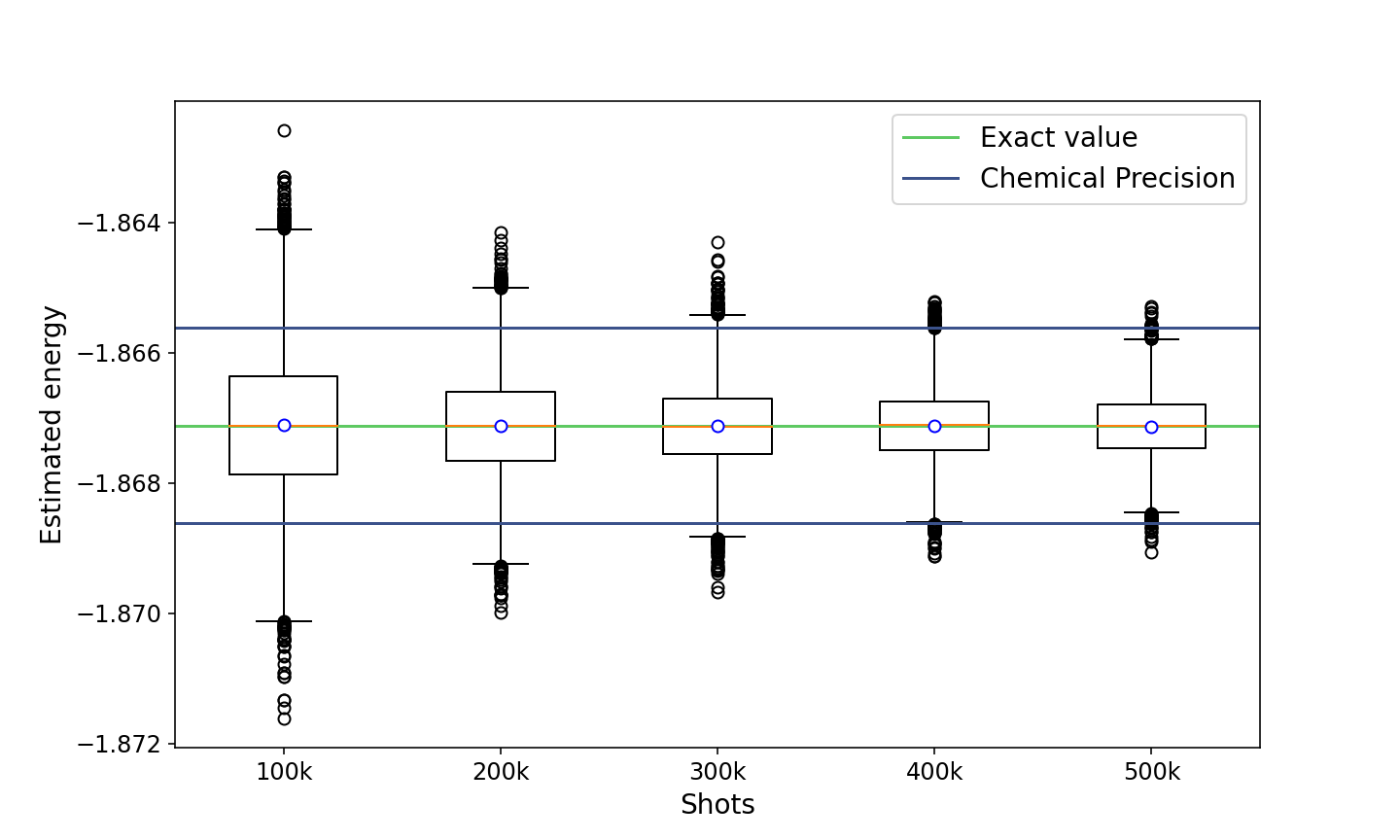}
     \caption{This boxplot represents 10000 energy values for different values of final estimation. The circle represents the mean of the distribution and the yellow line represents the median. It can be clearly seen here that even with the right eigenstate, one requires a large number of shots for the final estimation to get most of our results within chemical precision ($\pm0.0015$~Ha).}
     \label{fc1}
\end{figure}

\begin{figure}
    \centering
    \includegraphics[scale = 0.55]{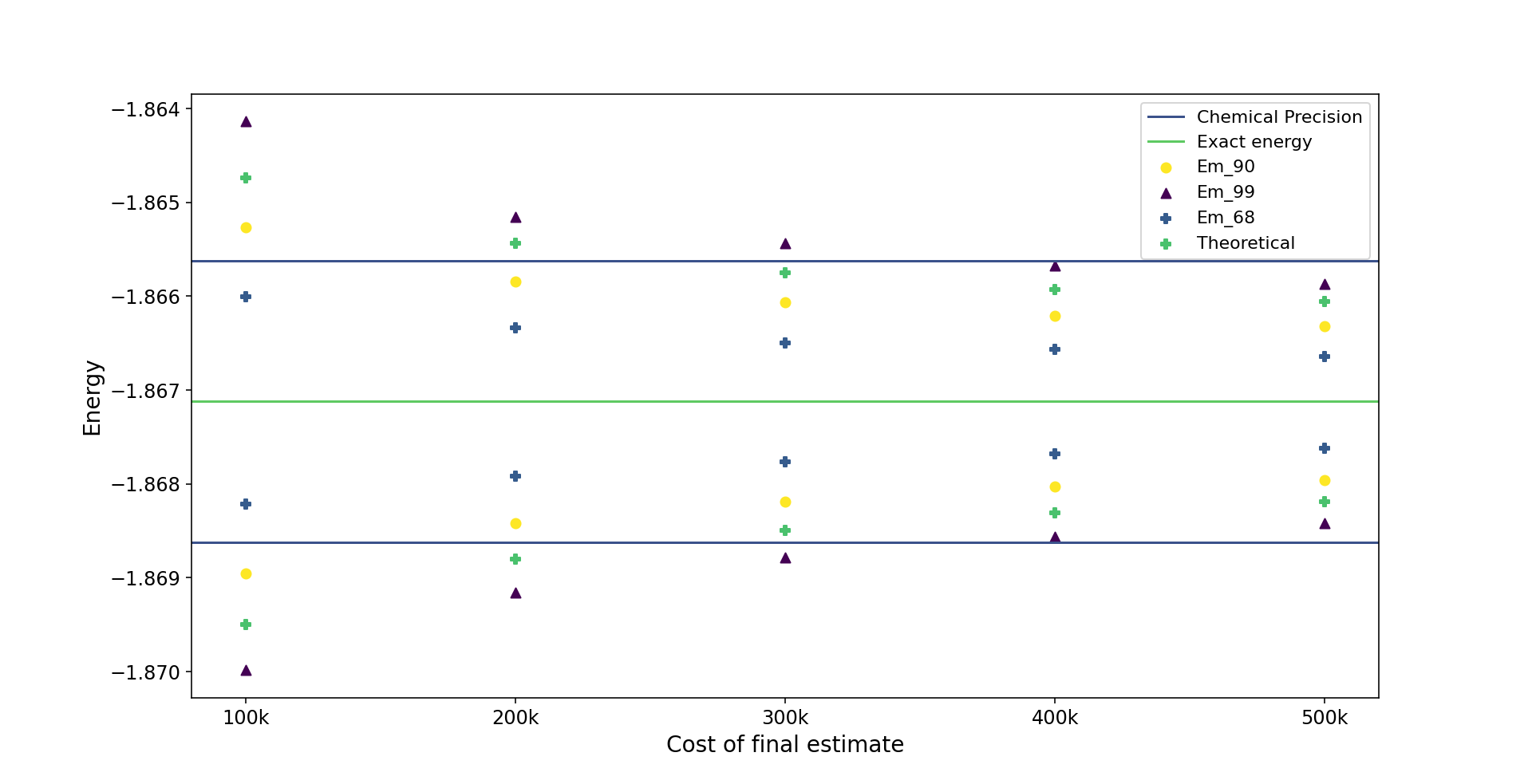}
    \caption{This plot shows a better comparison of theoretical and empirical estimates. $Em_{90}$ represents the end point of energy distribution after discarding extreme points so that 90\% of the energies are within this range. Similarly with $Em_{68}$ and $Em_{99}$. The theoretical estimate is based on the upper bound of the variance and should correspond to $Em_{68}$, however one can see that the bound is rather conservative and closed to about $95\%$. }
    \label{fig:bc}
\end{figure}

For chemical precision, the variation of energy with different final estimation is shown in Fig. \ref{fc1}.
To better comprehend how the theoretical estimate mentioned in the subsection "The cost of final estimation" matches with the empirical results shown in Fig. \ref{fc1}, we sorted the energy data points, discarded extreme points based on the confidence level selected, and then compared them to the theoretical estimate as shown in Eq. (\ref{h2}), which corresponds to a confidence level of 68\%. This is illustrated in Fig. \ref{fig:bc}.

It is worth to mention here that setting the number of shots in such a way that the final estimation will achieve the accuracy as desired might not be optimal in some scenarios. 
Naturally, the first and obvious one is when the total budget is comparable to the number of shots needed for estimation. Here one would significantly decrease the budget for the optimization itself and thus decrease the probability of obtaining a good result. 
In other words, we would know when we get the good result, but we never get it. In such a case it is better to search a bit more and risk a failure while increasing the probability of success. 
The other scenario is when the optimization procedure by itself is unreliable, thus resulting in a bad result in many cases even for a high-precision quantum subroutine. 
Here one will use a lot of repetitions and will have to reduce the final estimation costs accordingly.

\subsection*{Benchmarking of the optimization method} 

To be able to estimate the capabilities of the optimization method to deal with the optimization task depending on the available budget of shots, we will have to sample it. It is important to stress here that we will not set any internal parameters while sampling, like number of iterations, starting number of samples or the way the procedure changes the number of shots per measurement. 
The only criterion here will be the total number of shots used $n$ and the quality of the result measured in the probability to be within the desired precision. 

We model the method by an exponential function characterized by a set of three parameters 
\begin{equation}
        p_{s}=a(1-e^{-bn})+c, \label{fit}
\end{equation}
where $c$ is the ``guessing'' probability of success without any measurement (expected to be very close to zero for any realistic scenario), $a+c$ the probability of obtaining the desired result in the limit of infinite number of shots used (i.e. in the situation when the quantum subroutine would provide perfect outputs) and $b$ is the parameter expressing how the enhancement of the result depends on the number of shots used. 

To obtain such a parametrization, one needs to perform sampling of the procedure. This means running it repeatedly with different number of shots and evaluating the results. While the true minimum is not known, we expect the obtained results to have a mean around the physical value, i.e. there is no dominant local minima the optimization technique ends in. 
It is also important to stress that while the parameters $a,b$ and $c$ are dependent on the desired precision (the probability of success depends on the desired precision), all these values can be calculated from the same sampling data.  

\subsubsection*{Showcase of fitting on $H_2$ molecule}
\begin{figure}
    \centering
    \includegraphics[scale = 0.7]{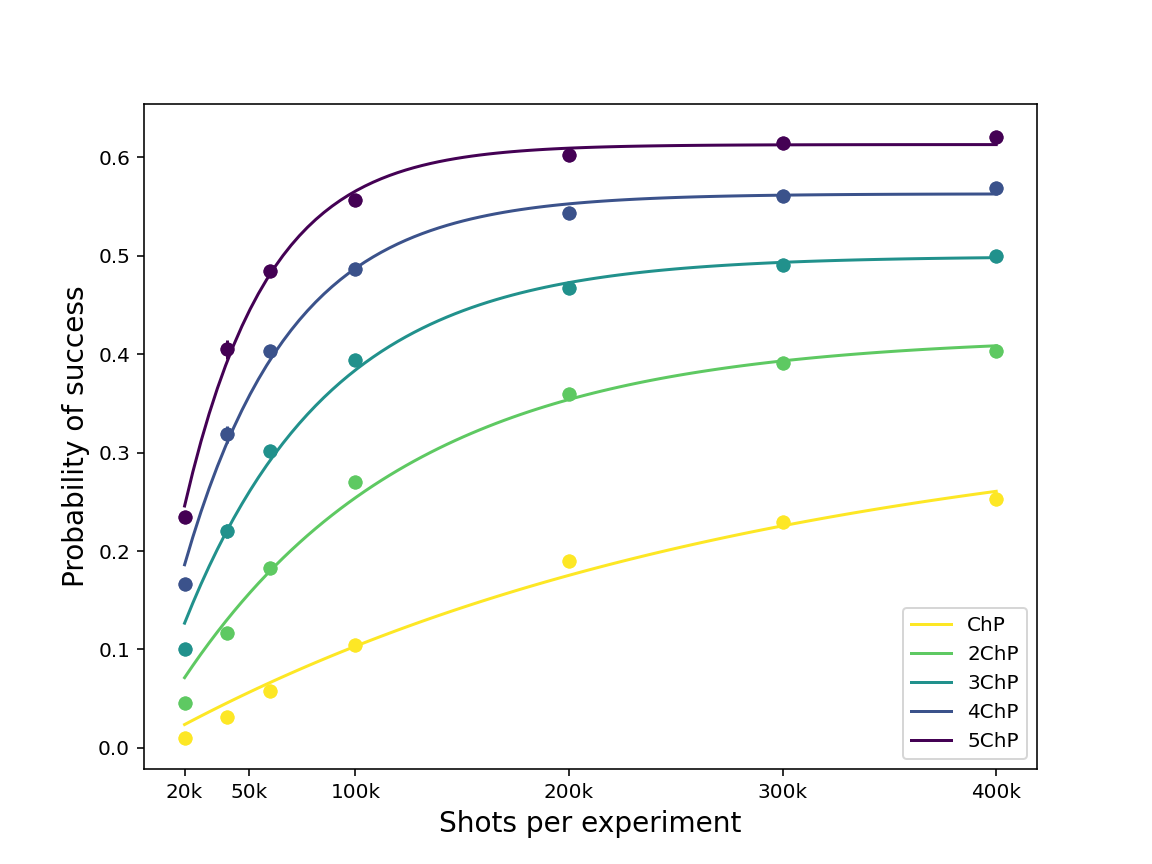}
    \caption{This is the pictorial representation of the probability of success with respect to shots per experiment which is shots per iteration $\times$ 100 (iterations). Differently colored lines represent different desired precision in multiples of chemical precission (ChP = $0.0015$~Ha).}
    \label{fig:fit}
\end{figure}

To evaluate our method of meta-optimization technique, we used $10000$ data points of energies to estimate probabilities of success for the energy to lie within the desired accuracy (expressed in multiples of the chemical precision (ChP)) by varying the shots per iteration, while fixing the \texttt{maxiter} (maximum number of iterations) to $100$. Results are depicted for different values of desired accuracy in Fig. \ref{fig:fit}. 

\begin{table}[]
    \centering
    \begin{tabular}{|c|c|c|c|c|}
    \hline
    Precision (Hartree)  & a & b & c & $a_{sv}$ \\
    \hline
    ChP      & 0.3416 & 3.60e-06 & 9.56e-11 & 0.3119 \\
    \hline
    2 $\times$ ChP  & 0.4185 & 9.35e-06 & 9.58e-18 & 0.4461 \\
    \hline
    3 $\times$ ChP & 0.4995 & 1.46e-05 & 3.71e-17 & 0.5313 \\
    \hline
    4 $\times$ ChP & 0.5629 & 2.01e-05
 & 1.8e-26 & 0.5913\\
    \hline
    5 $\times$ ChP & 0.613 & 2.56e-05 & 2.86e-17 & 0.6442\\
    \hline
    \end{tabular}
    \caption{This table shows the different values of fit parameters of the function $a(1-e^{-bn}) + c$ corresponding to different accuracy of energies, measured in chemical precision (ChP = $0.0015$~Ha). The last column $(a)_{sv}$ represents values obtained by using a state vector simulator instead of quantum measurements at each iteration in the optimization process. As one can see they correspond well to the obtain fitting parameters $a$, proving good extrapolation capabilities of the fit function. }
    \label{tab: Fit}
\end{table}

 In Table \ref{tab: Fit} we tabulate the resulting fit parameters for different values of accuracy. As one can see, the $c$ value is vanishing as expected, suggesting that random guessing of the state can hardly lead to a reasonable result. Value of $b$ expresses the inversed typical number of shots needed to start obtaining reasonable results. This ranges from about $40000$ for small accuracy to about $200000$ for large accuracy, what is a very reasonable result. 
 
 The $a$ value now expresses the limiting success probability of the optimization procedure. It ranges from about $30\%$ to $60\%$ depending on the desired accuracy, what might seem to be rather small. To validate this result independently, we did perform the same rounds of the optimization method, but using a state vector simulator instead of (simulation of) quantum computer, which virtually simulates infinite number of shots used. Results differ only by a few $\%$ compared to the fit parameters proving good extrapolation capabilities of the fit function.  

%We then used these values to fit the function of the form $a(1-e^{-bn}) + c$ and similarly obtained the best-fit parameters for different ranges of accuracy. These values are tabulated in Table. \ref{tab: Fit} . The fit function for chemical precision is illustrated in Fig \ref{fig:fit}.  
 
\subsection*{Repetitions}

Instead of investing the whole available budget into a single optimization that, as we have seen in the previous section, is likely to fail, it makes sense to repeat the procedure more times. 
On the other hand, very few shots have extremely low probability to succeed, so one might expect that there shall exist an optimum of number of repetitions of the whole procedure. 

If we repeat the experiment $r$ number of times, then the probability that at least one of the energies lies within desired range is given by 
    \begin{equation}
        P = 1 - (1-p_{s})^{r} = 1- (1-p_{s})^{B/n+m}, \label{eq2}
    \end{equation}
    where $B$ represents the total shot budget and $m$ represents the number of shots used for final estimation of the value of the observable. Note that $p_{s}$ is also a function of $n$. To get the best possible results, we have to maximize this total probability function
    \begin{equation}
        P_{max} =  \max_{n} (1-(1-p_{s})^{r}).  \label{eq3}
    \end{equation}
Instead of maximizing the above expression, one can minimize the following expression to locate the point of maxima
    \begin{equation}
         \min_{n} (1-p_{s})^{B/n+m} \equiv \min_{n} (1-p_{s})^{1/n+m}.
    \end{equation}
Once the optimal $n$ is found, one can calculate the number of repetitions ($r$) from $B,m$ and $n$ and repeat the experiment $r$ number of times to get the best possible outcomes.

\subsection*{Reliability}
Repeating the experiment $r$ times, we get $r$ different results. Then we have to select one of them that shall be presented as the final outcome of the meta-optimization technique. The only reasonable way to do that is to select the result that has the lowest energy stemming from the final estimation procedure. But while for large number of shots used for final estimation the outcoming result will be highly reliable, this is not the case for small budget anymore. In particular, if the precision of the final estimation will be comparable or lower than the desired precision defined for the meta-optimization technique, the single result provided as the outcome might not be the correct one. In other words, even if the optimization procedure in one of the $r$ runs did find a ``good'' result, the insufficient final estimation was not able to identify it correctly. 

To account for this aspect, we define a quantity called ``Reliable probability''. This is obtained by modifying the probability described in Eq. (\ref{eq2}) to incorporate the reliability of the energy value by multiplying it with the probability that the final estimate is good enough.
We first calculate the accuracy which can be obtained with a given budget for final estimation $E$ from the Eq. (\ref{eq-m})
 \begin{equation}
     \varepsilon = \sqrt{ \frac{\big(\sum_{i} \sqrt{\sum_{j} c_{ij}^{2})}\big)^{2}}{m}}.
 \end{equation}
  As this $\varepsilon$ corresponds to standard error, it lies at one $\sigma$ (standard deviation) away from the mean of our energy distribution. Thus we can then represent the desired precision ($d$) in terms of $\sigma$ to get the  
 Z-score (the number of standard deviations the value is above or below the mean value)
 \begin{equation}
     Z = \frac{d}{\sigma},
 \end{equation}
and convert it into the corresponding confidence level of the value to lie within the desired range ($\gamma$). Finally the probability ($P$) defined in Eq. (\ref{eq2}) is multiplied with this confidence level to get the reliable probability. In our case we define the final reliable probability as a multiplication of the maximal probability defined in Eq. (\ref{eq3}) with the reliability factor $\gamma$
 \begin{equation}
     P_{reliable} = \gamma P_{max}.
 \end{equation}

\subsection*{Results for $H_2$ molecule}
 
\begin{figure}
    \centering
    \includegraphics[scale = 0.6]{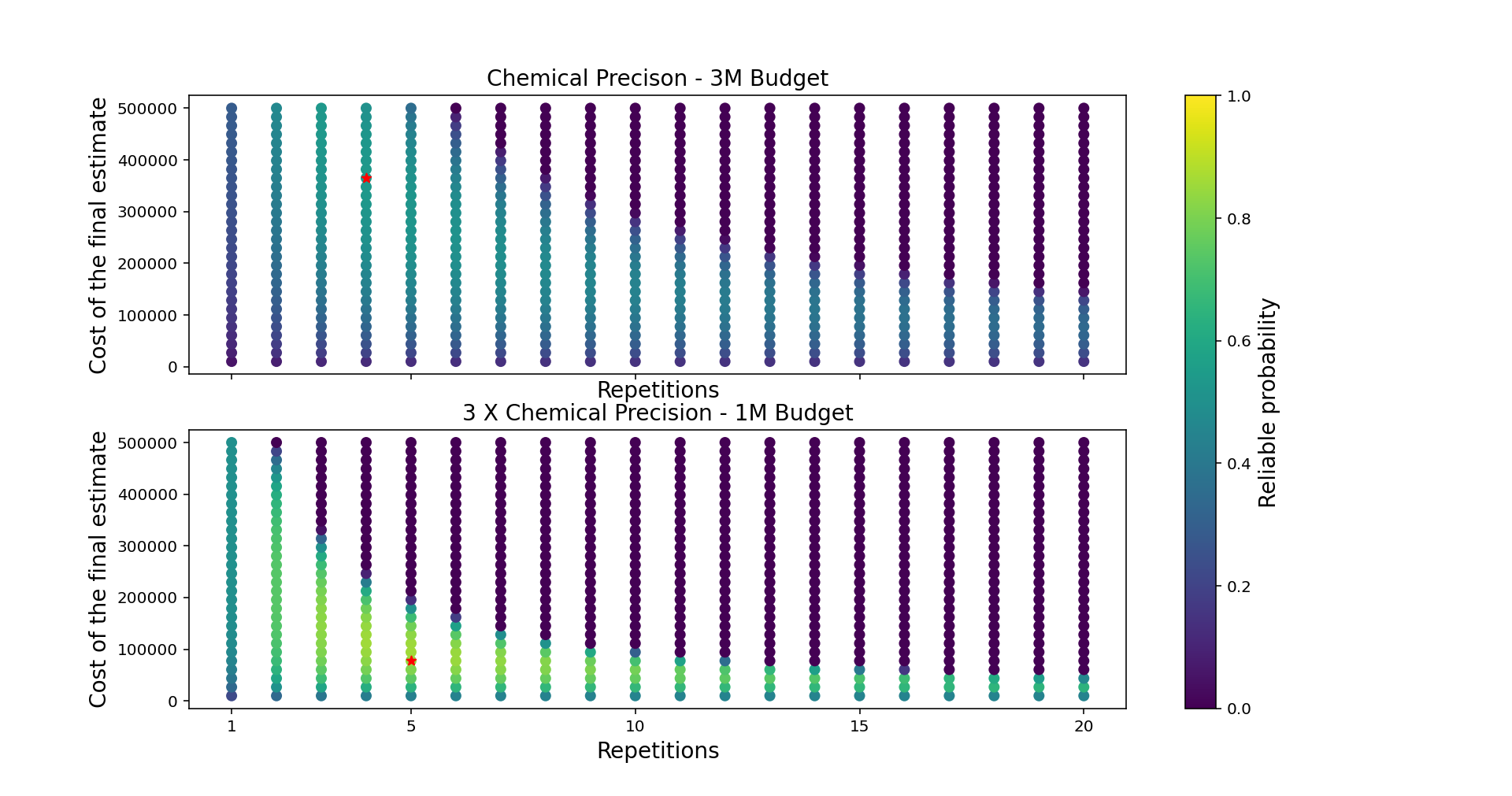}
    \caption{These plots represent the reliable probabilities to obtain a result with energy value within the desired range of accuracy for different values of final estimation and repetitions of the VQE for a fixed shot budget. The configuration which leads to maximum reliable probability is marked with a red asterisk. }
    \label{cmap_combi}
\end{figure}

 To showcase the working procedure of our technique, we provide a pictorial representation of reliable probabilities with varying repetitions and final estimation costs for different accuracy and budget (maximum number of shots that can be spent) in Fig. \ref{cmap_combi}. Here it is clearly visible that for a given budget there exists a clear optimum both for the number of repetitions of the classical optimizer and for the value of final evaluation. This can be understand as follows. 

 If the final evaluation cost is too large, not enough shots remain for the optimization itself. Thus the final result is evaluated correctly, but is likely out of the desired precision. On the other hand, if the final evaluation costs are too low, one of the results might be good, but we will likely choose the wrong one. 
Similarly, if the number of repetitions is too low, we might fail to get a good results simply due to the imperfectness of the underlying procedure itself. On the other hand, if the number of repetitions is too high, the budget per repetition is small and does not allow for obtaining a good result with reasonable probability. 

\begin{figure}
    \centering
    \includegraphics[scale = 0.7]{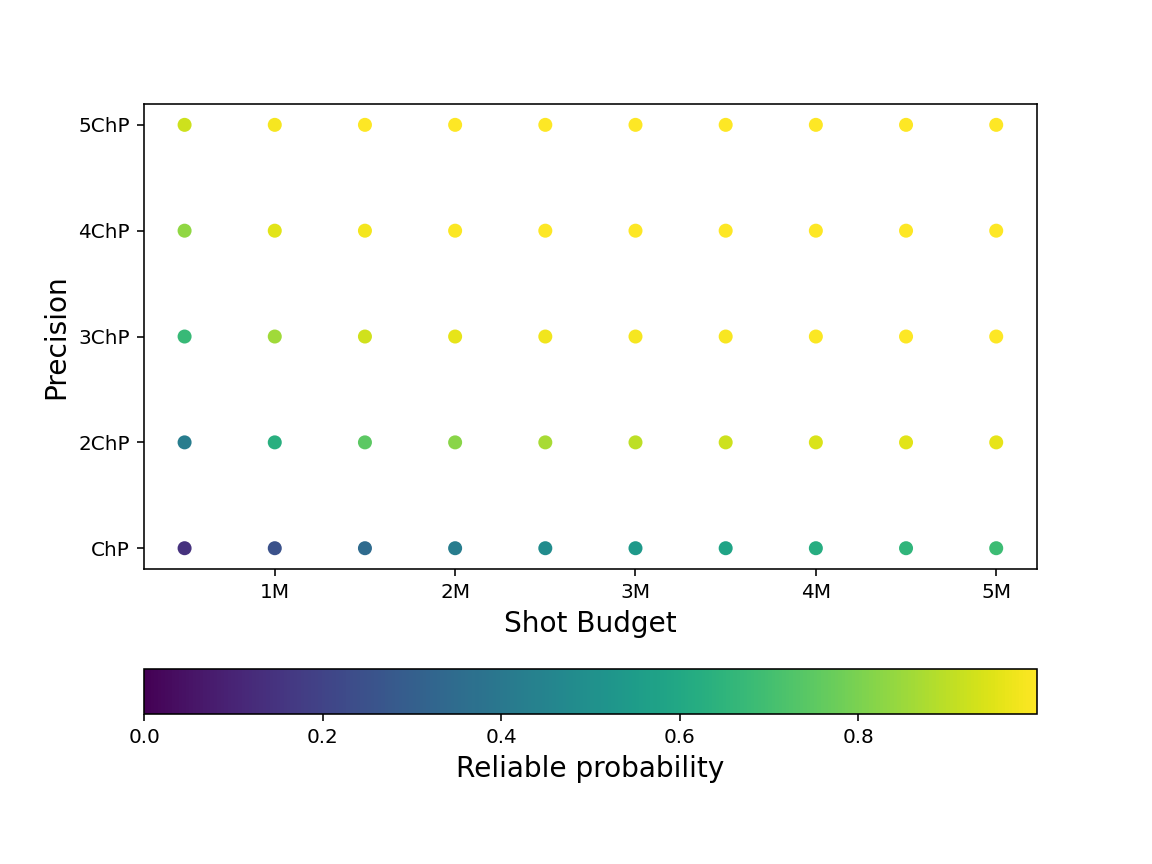}
    \caption{This figure illustrates the maximum reliable probabilities one can obtain for different shot budgets and accuracy. Every point in this figure is the red asterisk marked point in Fig. \ref{cmap_combi} for a different combination of accuracy and budget.}
    \label{fig:my_label}
\end{figure}

In the graph presented in Fig.\ref{fig:my_label} we show the final results of optimal reliable probability for a wide range of parameters. In essence, it aggregates the optima -- red asterisks depicted in graphs Fig. \ref{cmap_combi} -- for different input parameters. This graph showcases the capabilities of the meta optimization technique -- for a given budget of shots and desired precision, it gives us the information on the maximum probability of obtaining the desired result. For each point in the graph, though not depicted, we also obtain the information on how to achieve this probability, namely the optimal number of repetitions to be performed and the optimal distribution of shots between the optimization itself and the final evaluation. 

\section*{Discussion}

In this paper we have presented a meta-optimization technique that addresses two major obstacles in using NISQ quantum computers for variational tasks. Our goal is an efficient use of quantum devices by optimizing the number of times they are used during the run of the algorithm. 

Existing optimization techniques deal with two major issues. First, even if they are able to identify the optimal state relatively well, the result of the cost function is determined only with a very low accuracy. This is often circumvented by evaluating it in the end using classical means \cite{Kubler2020adaptiveoptimizer,rosalin,tamiya2022stochastic}, which is not feasible in scenarios with moderately large inputs. 
We address this problem by introducing the final evaluation step and optimizing the ratio of quantum resources used for comparison of the obtained results in multiple runs to the cost of optimization itself. It turns out that it might be far from negligible, i.e. to get a reliable result, one has to invest a good portion of quantum resources for its final evaluation. 

The other problem of existing techniques is that they are stochastic by their nature. This leads to the inevitable fact that in some cases it returns an incorrect result even if the quantum part works perfectly, i.e. with no noise and statistical deviations. 
Thus it makes sense to mitigate the risk of investing all quantum resources at hand into a single calculation. 
We presented a technique to benchmark the underlying optimization procedure and decide about the optimal number of repetitions. 
Interestingly, it turns out that this number is non-trivial (i.e. not one) for realistic scenarios. 

While we deployed our technique on a specific example of SPSA optimization on a hydrogen molecule, it can be applied on a very broad variety of variational quantum algorithms that are currently of a large interest to the quantum computation community. We believe that our meta-optimization can provide an easy to implement step towards practical use of quantum computers.

\section*{Methods}
%\guidelines{
%Topical subheadings are allowed. Authors must ensure that their Methods section includes adequate experimental and characterization data necessary for others in the field to reproduce their work.}

\begin{figure}[h]
     \centering
     \begin{subfigure}[h]{0.47\textwidth}
         \centering
         \includegraphics[width=\textwidth]{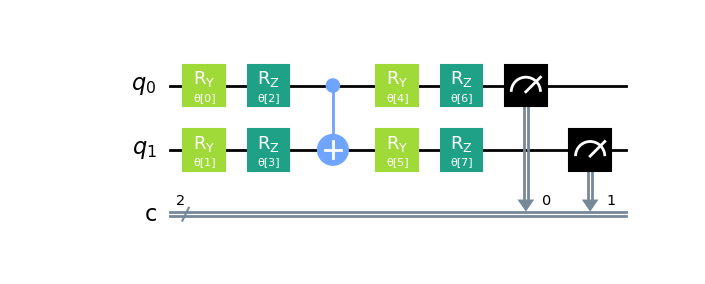}
         \caption{}
         \label{}
     \end{subfigure}
     \hfill
     \begin{subfigure}[h]{0.47\textwidth}
         \centering
         \includegraphics[width=\textwidth]{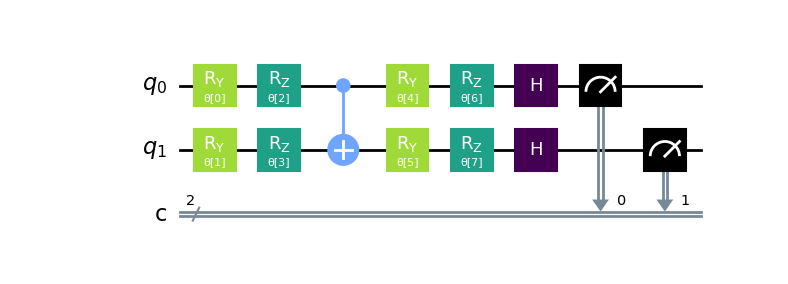}
         \caption{}
         \label{}
     \end{subfigure}
     \caption{ The qubits in the first circuit are measured in the computational \{$\ket{0}$,$\ket{1}$\} basis. To measure the qubits in the second circuit in \{$\ket{+}$,$\ket{-}$\} basis we applied Hadamard gates on both qubits in the end before the measurement. }
    \label{fig:preparation}
\end{figure}

For the quantum subroutine, we have used Ry-Rz ansatz to prepare the state from an initial zero state. This state was then measured in both $\{\ket{0},\ket{1}\}$ and $\{\ket{+},\ket{-}\}$ (the eigenstates of Pauli $X$ operator) basis to calculate the expectation value of the Hamiltonian. The procedure was defined by eight parameters $\theta[0]$ to $\theta[7]$ and circuits are shown in Fig. \ref{fig:preparation}.

\section*{Acknowledgments}

We acknowledge the support of VEGA project 2/0055/23 and GAMU project MUNI/G/1596/2019.
Further, we acknowledge the access to advanced services provided by the IBM Quantum Researchers Program. The views expressed are those of the authors, and do not reflect the official policy or position of IBM or the IBM Quantum team.  

\section*{Author contributions statement}
All authors contributed to the idea, design and writing of the manuscript. IA was responsible for the numerical simulations and producing graphs and figures.

\section*{Additional information}
\textbf{Competing interests:} Authors declare no compering interests;
\textbf{Data Availability:}
Data and programs used to derive the results presented in this paper are available from the corresponding author upon reasonable request.

\end{document}